\documentclass{article}
\usepackage{url}
\usepackage{graphicx}

\begin{document}
\title{Linking Quality Attributes and Constraints with Architectural Decisions}
\author{
David Ameller and Xavier Franch\\
{Universitat Polit\`ecnica de Catalunya}\\
{Barcelona, Spain}\\
{\{dameller, franch\}@essi.upc.edu}
}
\maketitle

\begin{abstract}
Quality attributes and constraints are among the main drivers of architectural decision making. The quality attributes are improved or damaged by the architectural decisions, while restrictions directly include or exclude parts of the architecture (for example, the logical components or technologies). We can determine the impact of a decision of architecture in software quality, or which parts of the architecture are affected by a constraint, but the difficult problem is whether we are respecting the quality requirements (requirements on quality attributes) and constraints with all the architectural decisions made. Currently, the common practice is that architects use their own experience to design architectures that meet the quality requirements and restrictions, but at the end, especially for the crucial decisions, the architect has to deal with complex trade-offs between quality attributes and juggle possible incompatibilities raised by the constraints. In this paper we present Quark, a computer-aided method to support architects in software architecture decision making.
\end{abstract}

\section{Introduction}
In the last decade, software architecture has become one of the most active research areas in software engineering. As a significant trend in this community, many researchers have stated that architectural decisions are the core of software architecture~\cite{Tyree2005}. Under this view, software architecture has evolved from a simple structural representation to a decision-centric viewpoint~\cite{Kruchten2009}. Along this way, several methodological approaches for architecture design and analysis had been proposed, e.g., ADD, TOGAF, ATAM, CBAM. These heavyweight methods offer a significant gain in the reliability of the architecture design process but require large-scale projects to achieve a balance between what the method offers and the effort that supposes for the architects to use it. This balance is hard to achieve when projects are low- or medium-scale. Lighter methods that apply for not so large projects are required~\cite{Tang2006}.

In this context, we present Quark (Quality in Architectural Knowledge), a computer-aided method to support architects in software architecture decision making. This method is being implemented in a tool, called ArchiTech~\cite{Ameller2011b} that acts as a proof of concept for the Quark method. ArchiTech is already capable to manage AK and to reason about it in a similar way as Quark describes.

Quark builds upon the reuse of Architectural Knowledge (AK) represented as an ontology called Arteon~\cite{Ameller2011}. Arteon includes knowledge about  architectural decisions, including their rationale and their link to quality attributes and constraints. Among other alternatives, we have chosen to use ontologies to represent the AK because ontologies have been successfully used for knowledge representation in other domains (e.g., software engineering, artificial intelligence, semantic web, biomedicine, etc.) and they offer other advantages such as reasoning and learning techniques ready to be applied (e.g., we could add new decisions using case-based reasoning techniques). The AK acquisition is not addressed in this paper, although we recognize that resolve the issues related to the acquisition of AK are essential to make this method useful for practitioners.

The rest of this paper is divided in the following sections: the Quark method (Section~\ref{quark_section}), an example of use of Quark (Section~\ref{example_section}), the related work (Section~\ref{relwork_section}), and finally conclusions and future work (Section~\ref{conc_and_fw_section}).

\section{The Quark Method} \label {quark_section}
\begin{figure}[!t]
\centering
\includegraphics[width=\columnwidth]{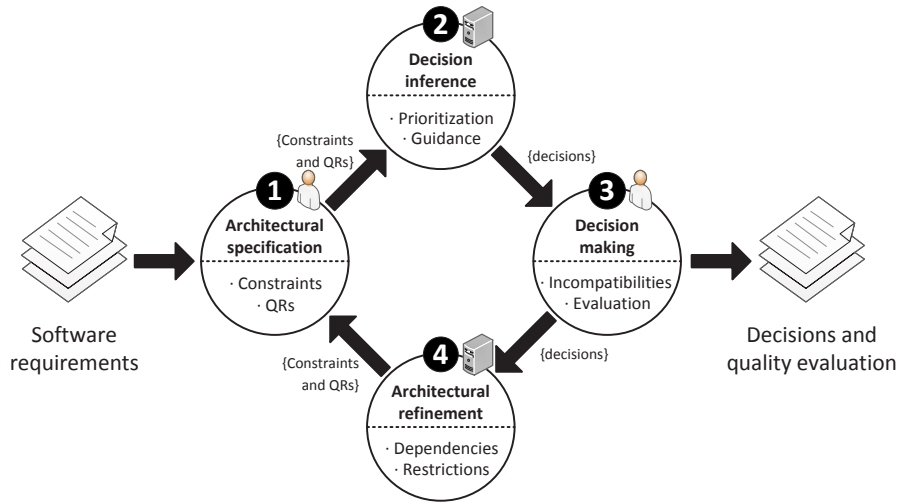}
\caption{Quark overview.}
\label{quark_fig}
\end{figure}

Quark aims at providing means to facilitate and making more reliable architects' decisions with regard to quality attributes. The method starts with the software requirements, and ends with a set of architectural decisions and the overall evaluation of the quality attributes (see Figure~\ref{quark_fig}). The Quark method delivers an iterative process divided in four activities: first, specification of the Quality Requirements (QR) and the imposed constraints; second, inference of architectural decisions according to existing AK; third, decision making; and fourth, architectural refinement.

The design of Quark has been driven from a critical observation gathered from two empirical studies we have recently conducted\footnote{Not available as publication yet.}: software architects may be receptive to new design methods as far as they still keep the control on the final decisions. In other words, architectural decision making should not be an automatic process, but just computer-aided, therefore still architect-driven. As a consequence, the architect plays the central role in Quark. In particular, there are two tasks where the architect role takes special relevance. First, in activity 1, the architect has to define the QR and constraints relevant to the software architecture. Second, in activity 3, the architect has to choose the architectural decisions and then decide if the decisions made are sufficient to end the process. In the same direction, Quark is intended to provide guidance but not to control the decision making process, in this sense Quark will notify about possible incompatibilities and possible actions to resolve them, but the method does not require resolving detected incompatibilities to continue, the architect has always the last word.

In the following subsections we describe each activity.

\subsection{Architectural Specification} \label{quark_specification_subsection}
\begin{figure}[!t]
\centering
\includegraphics[width=\columnwidth]{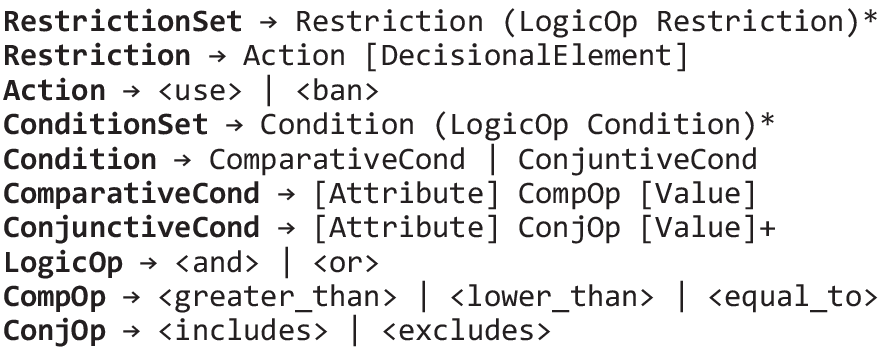}
\caption{CFG to formalize constraints.}
\label{CFG_fig}
\end{figure}

In the first activity, the architect specifies manually the QRs and constraints that will drive the architecture decision making. These QRs and constraints are the architect interpretation of the software requirements that are relevant for the architecture design. For example, a QR could be ``performance should be high'' (in other words, more a goal than a requirement) or something more concrete such as ``loan processing response time should not be higher than two seconds 95\% of the times''. Examples of constraints could be ``the database management system (DBMS) must be MySQL 5'', and ``the architectural style must be Service-Oriented Architecture (SOA)''. In order to be computer-aided this specification should be formalized. We provide an example expressed as a Context Free Grammar (CFG) (see Figure~\ref{CFG_fig}). For simplification, we included extra notation in the CFG: \emph{[concept]} mean one valid instance of the concept in the Arteon ontology~\cite{Ameller2011} and \emph{$<$symbol$>$} mean a terminal symbol.

Due to the Quark's iterative nature, the specification of these QRs and constraints does not need to be complete, which makes this method less heavy. The architect could start from a very short specification and then grow the architecture in each refinement or start from a complete specification and see if it matches the expected QRs, and then if necessary refine it.

\subsection{Decision Inference} \label{quark_inference_subsection}
In the second activity, the Quark method uses the AK available to generate a list of decisions. Since the expected amount of decisions is large, the decisions should be ordered by priority criterion (e.g., the decisions that satisfy more constraints  and better comply with the stated QRs are top priority). This criterion will be used in an AI local search algorithm (e.g., simulated annealing). The criterion can be tuned by the architect.

Another observation from our empirical studies is that the decisions generated need to be descriptive. For example, we could describe ``data replication'' decision as follows: ``the decision is offered because there is a requirement about having high performance'', ``by making this decision, the overall performance will increase but will affect negatively to the maintenance, and can damage the accuracy'', ``also, by selecting this decision, the used DBMS is required to be able to operate with data replication.'' All this information can be obtained from the ontology~\cite{Ameller2011}. More complete templates to describe decisions are available in the literature (e.g.,~\cite{Tyree2005}).

\subsection{Decision Making} \label{quark_decisionmaking_subsection}
In the third activity, the architect decides what decisions are to be applied from the ones generated in the previous activity. When the architect makes a decision, some issues may rise. For example, there could be incompatibilities with previous decisions (e.g., we are selecting the ``data replication'' decision, but we already selected a DBMS that does not support data replication), or there could be some QRs that do not hold by the decisions made (e.g., the decisions indicate that maintainability will be damaged while there is a QR that says that maintainability is very important for this project). These issues may be detected using constraint satisfaction algorithms.

As said before, the architect will be informed about which decisions are the most conflictive, but s/he will decide if the actual decisions are adequate or not (e.g., there may be external reasons, beyond requirement satisfaction, that have higher priority). Also, at this point the architect has to decide if s/he wants to end the decision making process or make a new iteration (see refinement activity).

\subsection{Architectural Refinement} \label{quark_refinement_subsection}
In the fourth activity, the objective is to make actions that will resolve the detected issues. We identified three possible outcomes from the decision making activity: incompatibilities, dependencies, and suggestions for QRs. Incompatibilities were explained in activity 3. Dependencies occur when some decision requires the existence of other elements in the architecture. For example, when the architect decides to use SOA, several related decisions are needed: service implementation (e.g., SOAP, REST), service granularity (e.g., service composition, single service), etc. Suggestions for QRs are inferred because some quality attribute is having special relevance due to the selected decisions, e.g., most of the decisions have positive impact on security. In that case we may suggest to the architect to consider including a QR about security. This help architects to make QRs explicit.

The three possible outcomes are to be included in activity~1 for the next iteration, incompatibilities and dependencies may be translated into constraints, suggestions will add new QRs. Once in activity~1, the architect will decide which of these outcomes will be included in the next iteration and at this point, the architect can add or modify the previous iteration constraints and QRs. For example, the architect may have noticed in the last iteration that one QR was very difficult to meet and decide to soften it.

\section{Example} \label{example_section}
A complete architectural decision making process is too long to be included, the present example will focus only in one decision, the DBMS selection. Although the example is centered in a decision about technologies, a similar process could be applied to a decision on architectural patterns.

Following the Quark method, before starting the architect should revise the software requirements document of the project and identify the ones that are relevant to the architecture. For this example, the relevant requirements are: (R1) ``the software system shall keep the information about clients and providers'', (R2) ``the software system shall be developed using OSS whenever possible'', and (R3) ``the software system shall have high reliability''.

\subsection{Specification Activity} \label{example_activity1_subsection}
Once software requirements are selected, the software architect should interpret them and provide the resulting QRs and constraints. From the R1 the architect may deduce that the project is an information system, so a DBMS will be required. From R2 the architect may include a constraint on the technologies used to be OSS. From the R3 the architect may include constraints to have backup facilities, and a QR for reliability. Using the formalization presented in Figure~\ref{CFG_fig} the specification will be: \emph{Use} DBMS, ``License'' \emph{includes} {``GPL'', ``LGPL'', ``BSD'', etc.}, ``Backup facility'' \emph{equal} ``yes'', and ``Reliability'' \emph{greater than}  ``average''.

\subsection{Decision Inference Activity} \label{example_activity2_subsection}
In this example we use as AK the information published in~\cite{POJ_comparisonDBMS}, and the prioritization criterion mentioned in Section~\ref{quark_inference_subsection}. The following possible decisions are generated:

\begin{enumerate}
\item The decision to use MySQL 5 is offered because it is OSS. There is no information available about backup facilities in MySQL. MySQL is preferred because it supports more OSS technologies. Using MySQL has neutral impact in reliability because ACID compliance depends on the configuration.
\item The decision to use PostgreSQL 8.3 is offered because it is OSS. There is no information available about back-up facilities in PostgreSQL. There are few OSS technologies with support for PostgreSQL. Using Postgre-SQL improves reliability because it is ACID compliant.
\item The decision to use SQL Server 2005 is offered because it satisfies the backup facility condition. SQL Server is not OSS. There are few OSS technologies with support for SQL Server. SQL Server will require a Windows operating system. Using SQL server improves reliability because it is ACID compliant.
\end{enumerate}

\subsection{Decision Making Activity} \label{example_activity3_subsection}
In the decision making activity, the architect, for example, will decide to use MySQL 5 (the decision with higher priority) as the implementing technology for the DBMS component. But as said before in this paper, the architect may prefer to use PostgreSQL, even it is not the top decision, because s/he is more familiar to it (or any other reason). The important point is that the architect is able to make informed decisions, and, eventually, new decisions that were unknown to her/him are taken into consideration.

\subsection{Architectural Refinement Activity} \label{example_activity4_subsection}
After the decision making activity the architectural design will continue with new iterations, where the decision to use MySQL will impact, for example, in the selection of other technologies that are compatible with MySQL. This information will appear during the refinement activity as dependencies and incompatibilities.

\section{Related Work} \label{relwork_section}
Hofmeister et al.~\cite{Hofmeister2007} compared five software architecture design methods and came out with a general model of architectural design method composed of three activities: architectural analysis, architectural synthesis, and architectural evaluation. The Quark activities are not very different of these ones except for two differences: first, the general approach does not consider iterative methods, which is why we have an extra activity in our method, and second, Hofmeister's general approach deals with complete architectural solutions, while Quark works at decisional level. In our empirical study architects said that they will not trust a computer-aided method that generates full architectural solutions without their intervention.

Tang et al.~\cite{Tang2009} presented AREL, a UML-based model to explore the architectural design reasoning. With this model they are able to represent the decision making, and it helps to share a common understanding of the rationale behind each decision by having well documented decisions. They also say that with this model they help to identify trade-offs, issues, constraints, etc. In Quark, we do a similar task, but thanks to the use of ontologies to manage AK and the formalization of the QR and constraints we are able surface these trade-offs, issues, constraints in a semi-automatic way with computer-aided mechanisms. Also, the decisions generated by Quark are well documented (as explained in Section~\ref{quark_inference_subsection}). There is another difference between approaches, while AREL is thought to manage the decisions made, Quark is thought to propose decisions that the architect may want to make.

There is a documentation framework for architecture decisions proposed by van Heesch et al.~\cite{Heesch2012}. It consists of four viewpoint definitions using the conventions of ISO/IEC/IEEE 42010: a decision detail viewpoint, a decision relationship viewpoint, a decision chronology viewpoint, and a decision stakeholder involvement viewpoint. The output of Quark is a set of decisions, similar to the first viewpoint. The other three viewpoints are not supported, because they refer to architectural aspects that are not representable in the ontology in which Quark is built upon. We do not discard to include these aspects in the future.

\section{Conclusions and Future Work} \label{conc_and_fw_section}
In this paper we have presented the Quark method, a method to support architects in software architecture decision making. The key features of this method are:

\emph{Computer-aided}: Quark facilitates some parts of the decision making process, principally: the generation of a priority  list of decisions and the detection of incompatibilities.

\emph{Lightweight}: Quark guides the architect in the decision making by giving suggestions and indications, with this we 
expect to reduce the time to design the architecture.

\emph{Flexibility}: Quark may start just with few QRs and constraints or with a full specification.

\emph{Ontology-based}: Quark is built upon an ontology. Ontologies are well known in AI, and there are techniques ready to be applied for the reasoning with knowledge that is represented using ontologies.

\emph{Complement architects' knowledge}: The reuse of AK may help to complement architects' knowledge, e.g., the architect may not know all the effects of using a technology or an architectural pattern, it may need of other components to work correctly or it may have unexpected effects on the evaluation of some quality attributes.

Future actions with regard to this method include: finish the ArchiTech tool~\cite{Ameller2011b} that serves as a proof of concept of this method. Next, use the method in a real project as a validation strategy (we are in collaboration with one software company, Everis, that may fit to this end), and then include it in our envisioned framework using Model-Driven Development techniques~\cite{Ameller2010}.

There are two limitations in Quark that need to be adjusted: first, Quark is thought to be used at the beginning of a project as the initial decision making, it would be more useful if the method could be used during the whole development time, and second, Quark only deals with the structural parts of the architecture (e.g., patterns, styles, technologies, etc.) other aspects need to be supported as mentioned in Section~\ref{relwork_section}.

Finally, issues related to AK acquisition, AK maintenance, and AK reusability should have more attention from the architecture community because these topics are crucial to make this and other approaches suitable for practitioners.

\section*{Acknowledgments}
This work has been partially supported by the Spanish MICINN project TIN2010-19130-C02-02.  Thanks to Oriol Collell for his comments.

\bibliographystyle{abbrv}
\bibliography{CoRR}

\end{document}